\documentclass[fleqn,12pt,twoside]{article}
\usepackage{espcrc1}

\usepackage{graphicx}

\title{Spectroscopy in ep collisions at HERA}

\author{D. Ozerov\address[ITEP]{A.I. Alikhanov 
    Institute for Theoretical and Experimental Physics, \\
    Bolshaya Cheremushkinskaya 25, 117218 Moscow, Russia,\\
    and Deutsches Elektronen-Synchrotron,
    Notkestrasse 85, 22607 Hamburg, Germany}
  (on behalf of the H1 and ZEUS Collaborations)
}
       
\begin{document}

\maketitle

\begin{abstract}
Recent spectroscopy studies performed in high energy ep collisions by the H1
and ZEUS Collaborations are described. The status of the searches for 
the strange pentaquark, the {$\theta^+$}, is presented as well as results
on searches for further exotic states such as the $\Sigma^{--}$. The evidence
for the observation of a pentaquark state containing the charmed quark, 
the $\theta_c$,
is discussed and the result is given on searches for glueball candidates 
at HERA.
\end{abstract}

\vspace{4mm}

\begin{figure}[htb]
\begin{minipage}[t]{80mm}
\includegraphics*[scale=0.46]{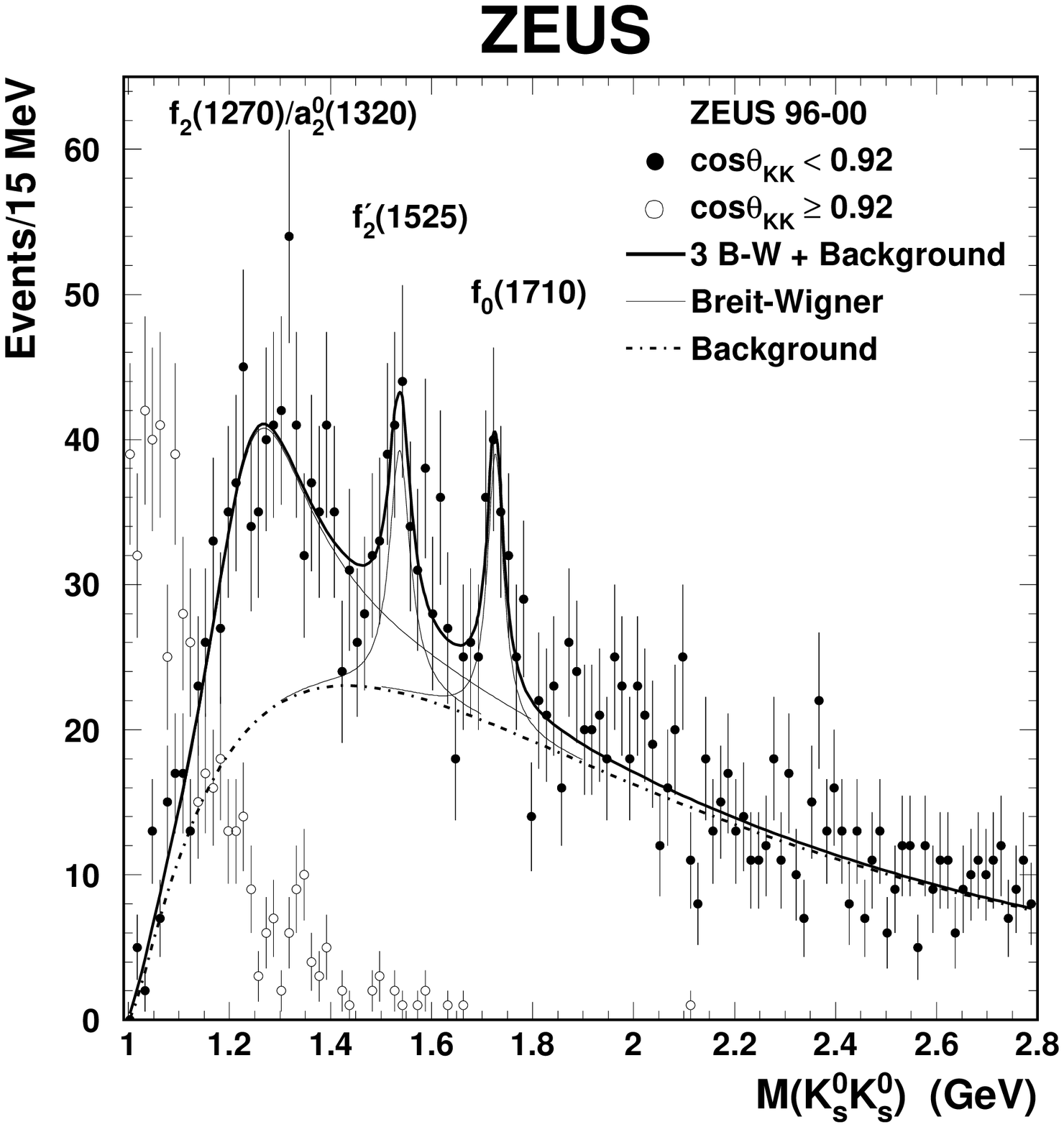}
\caption{The $K^0_sK^0_s$ invariant-mass spectrum for $K^0_s$ pair candidates
with $cos\theta_{K^0_sK^0_s} < 0.92$ (filled circles). The thick solid line
is the result of a fit using three Breit-Wigner (thin solid lines) and
a background function (dotted-dashed line). The $K^0_s$ pair candidates that
fail the $cos\theta_{K^0_sK^0_s} < 0.92$ cut are also shown (open circles).}
\label{fig:k0k0}
\vspace{-7mm}
\end{minipage} 
\hspace{\fill}
\begin{minipage}[t]{76mm}
\includegraphics*[scale=0.37]{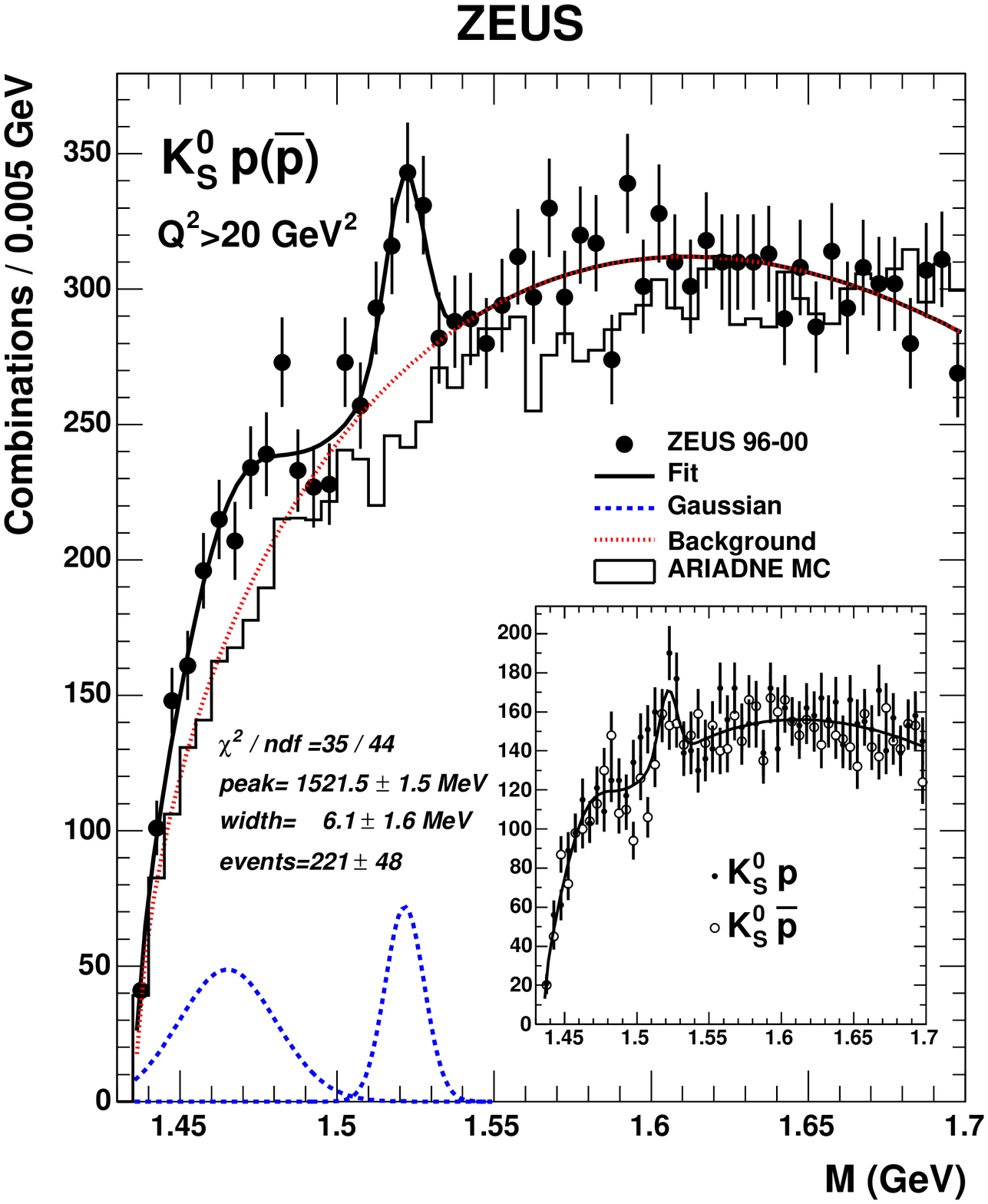}
\caption{Invariant mass for the $K^0_sp(\bar{p})$ resonance. 
  The histogram depicts
  the predictions from the Monte Carlo simulation and is normalized to the
  data in the region above 1650 MeV. The inset shows the 
  measurements for the $K^0_sp$ (black dots) and $K^0_s\bar{p}$ 
  (open circles) candidates separately.}
\label{fig:zeus_spq}
\vspace{-7mm}
\end{minipage}
\end{figure}

During the first stage of operation of the unique ep collider HERA,
the two Collaborations H1 and ZEUS collected data samples corresponding
to more than 100 $pb^{-1}$ each. Good tracking detectors
allow the identification of various resonances via their charged 
particle decay modes. One study was performed by the 
ZEUS Collaboration using events
with at least two well identified $K^0_s$, detected via its decay to 
two charged pions.
Such a $K^0_sK^0_s$ system is of particular interest since it is
expected to couple to scalar and tensor glueballs~\cite{glue_k0k0}.
Lattice QCD calculations predict~\cite{glue_lattice} the existence of 
a scalar glueball with a mass of $1730\pm100$ MeV,
close to the observed $f_0$(1710) state. 
The details of the event selection and analysis can be found 
in \cite{zeus_k0k0}. Figure~\ref{fig:k0k0} shows the measured $K^0_sK^0_s$ 
invariant mass spectrum. In the region above 1500 MeV, two narrow resonances
can be seen. The lower mass state has a fitted mass of $1537^{+9}_{-8}$ MeV
and a width of $50^{+34}_{-22}$ MeV, the higher mass state has a fitted mass
of $1726\pm7$ MeV and a width of $38^{+20}_{-14}$ MeV. The first state is in
a good agreement with the $f^{'}_2$(1525) state, while the second is consistent
with the glueball candidate $f_0$(1710). It is interesting to note
that the selected $K^0_s$-pair originated predominantly from the region
where sizeable initial state gluon radiation may be expected.

\begin{figure}[htb]
\begin{minipage}[t]{70mm}
\includegraphics*[scale=0.4]{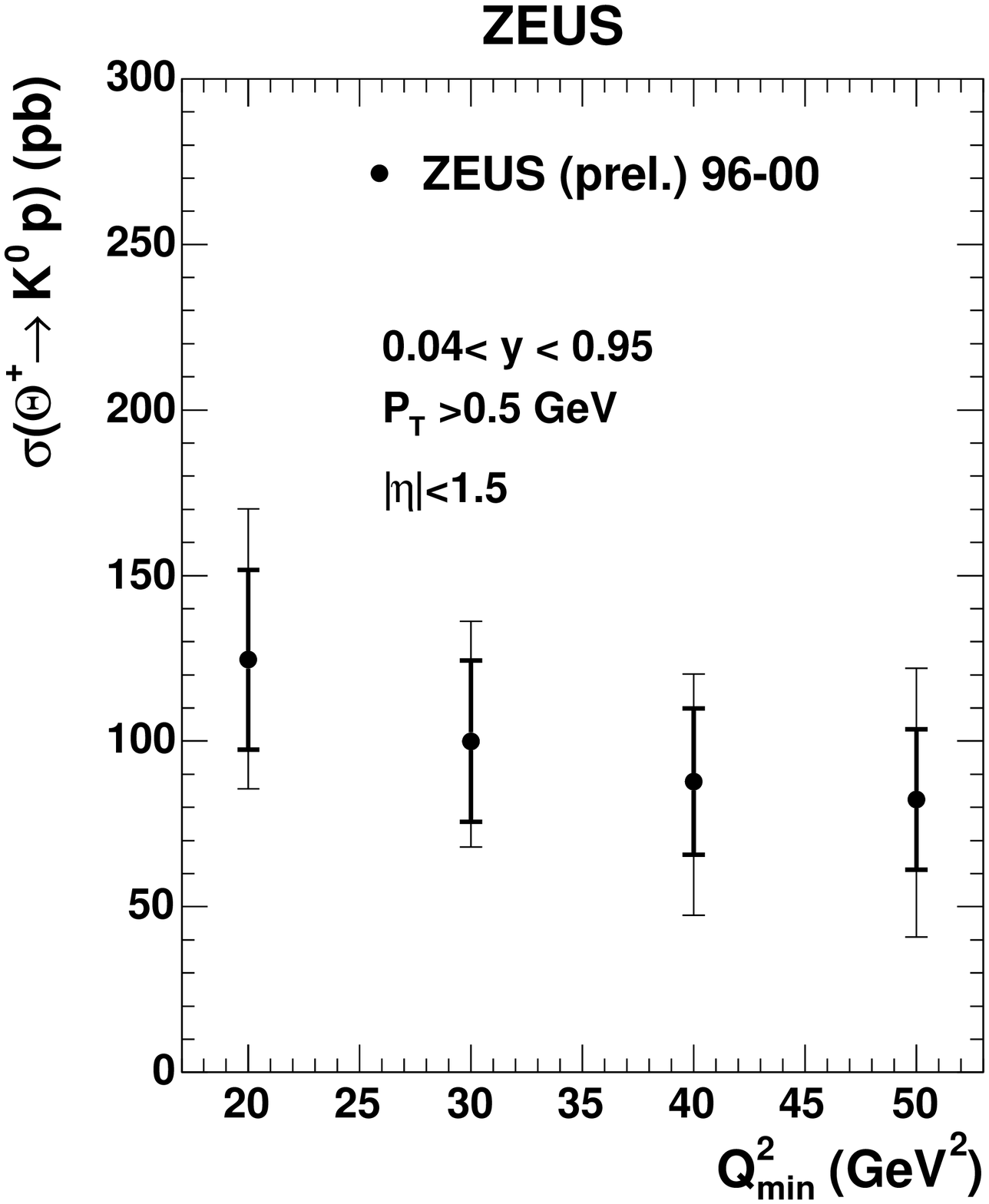}
\caption{Visible cross sections for the $\theta^{\pm}$ baryon decaying
  to $K^0_sp(\bar{p})$ as a function of $Q^2_{min}$.}
\vspace{-6mm}
\label{fig:zeus_spq1}
\end{minipage}
%
\hspace{5mm}
\begin{minipage}[t]{70mm}
\includegraphics*[scale=0.4]{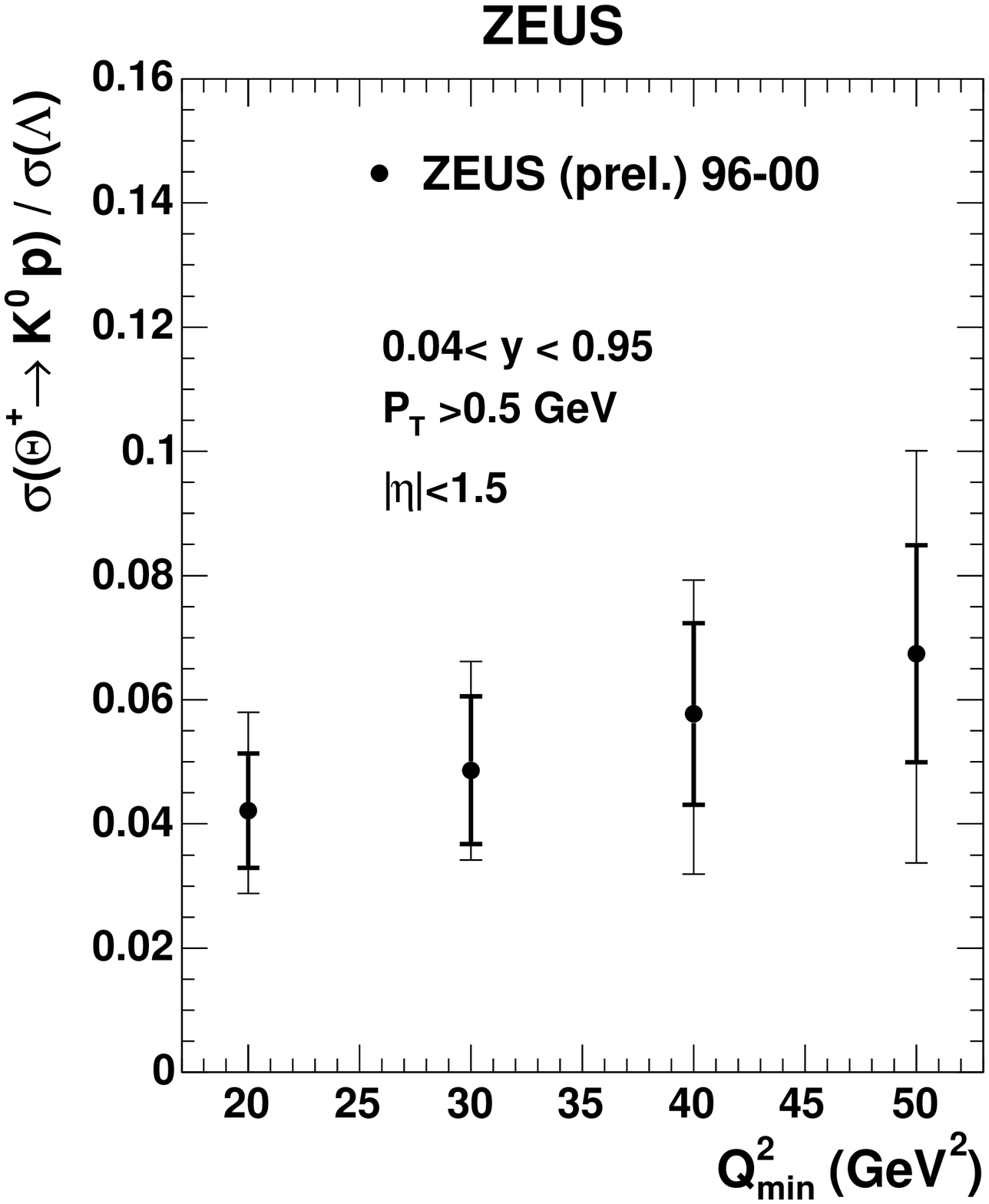}
\caption{Cross sections ratio \newline 
  $R = \sigma(\theta^{+} \to K^0_sp)/\sigma(\Lambda)$
  as a function of $Q^2_{min}$.}
\vspace{-6mm}
\label{fig:zeus_spq2}
\end{minipage}
\end{figure}

The hot topic in spectroscopy today is the discussion of the possible
existence of a pentaquark state with extremely narrow width and low mass.
Such states were predicted in \cite{pq_dpp}. The ZEUS Collaboration 
analysed the system $K^0_sp(\bar{p})$, where $K^0_s$ was again
identified via its decay to charged pions and the (anti)proton by
measuring the charged particle ionization. Details can be found in 
\cite{zeus_spq}. Figure~\ref{fig:zeus_spq} shows the $K^0_sp(\bar{p})$ 
invariant mass for $Q^2 > 20$ GeV$^2$, where $Q^2$ is the
squared momentum transfer between lepton and nucleon. The fit of two Gaussians
on top of a continuous background is shown on the picture. The
first Gaussian describes the narrow peak at 1522 MeV, while
the second Gaussian, near 1470 MeV, represents the complicated 
background shape in this region, which may be due to the broad resonance
$\Sigma$(1480).
The peak position obtained from the fit is 
$1521.5\pm1.5(stat.)^{+2.8}_{-1.7}(syst.)$ MeV. The Gaussian width was 
found to be $6.1\pm1.6$ MeV, compatible with the experimental 
resolution of 2 MeV. The fit of the Breit-Wigner, convoluted with a
Gaussian with a width fixed to the experimental resolution, gives 
a signal BW-width of $8\pm4(stat.)$ MeV. The number of events obtained 
from the fit is $221\pm48$, which corresponds to the statistical 
significance of $4.6\sigma$. The fit of the $K^0_s{\bar{p}}$ combinations
gives $96\pm34$ events which agrees with the signal extracted from
the $K^0_sp$ channel. If the $K^0_sp$ signal corresponds to the $\theta^{+}$
pentaquark observed by several other experiments, this measurement provides the
first evidence for the antiparticle $\theta^{-}$ with the quark
content of $\bar{u}\bar{u}\bar{d}\bar{d}s$. Figure~\ref{fig:zeus_spq1} shows 
the $\theta$ production cross section for $Q^2 > Q^2_{min}$, 
measured in the kinematic
region $0.04<y<0.95$, $P_T(\theta)>0.5$ GeV and $|\eta(\theta)|<1.5$
(where $y$ is the lepton inelasticity), as a function of $Q^2_{min}$.
The cross section ratio $R$ of $\theta$ to the well-known $\Lambda$(1116) 
baryonic state
was measured to be $R=(4.2\pm0.9^{+1.2}_{-0.9})\%$ for
$Q^2 > 20$ GeV$^2$.
Figure~\ref{fig:zeus_spq2} shows $R$ for several $Q^2_{min}$ values.

The $pp$ fixed target NA49 Collaboration reported~\cite{na49_sspq} observation
of the $\Xi$ multiplet pentaquark candidates $\Xi^{--}_{3/2}$ and
$\Xi^0_{3/2}$, predicted in \cite{pq_dpp} together with 
the $\theta^{+}$ pentaquark.
The narrow resonances were observed in $\Xi\pi$ combinations at masses 
$\approx 1862$ MeV. ZEUS performed~\cite{zeus_sspq} a search for such 
states.
No evidence of a signal at about 1860 MeV was found in any 
$\Xi\pi$ combination, while a
clean $\Xi^0(1530) (\to \Xi\pi$) is seen with a significance 
of $\approx 4.8\sigma$.
The discrepancy in the results of the two experiments can be attributed to 
the fact that ZEUS made the search
in the central rapidity region, while NA49 covers the forward 
rapidity region.

\begin{figure}[htb]
\begin{minipage}[t]{75mm}
\includegraphics*[scale=0.60]{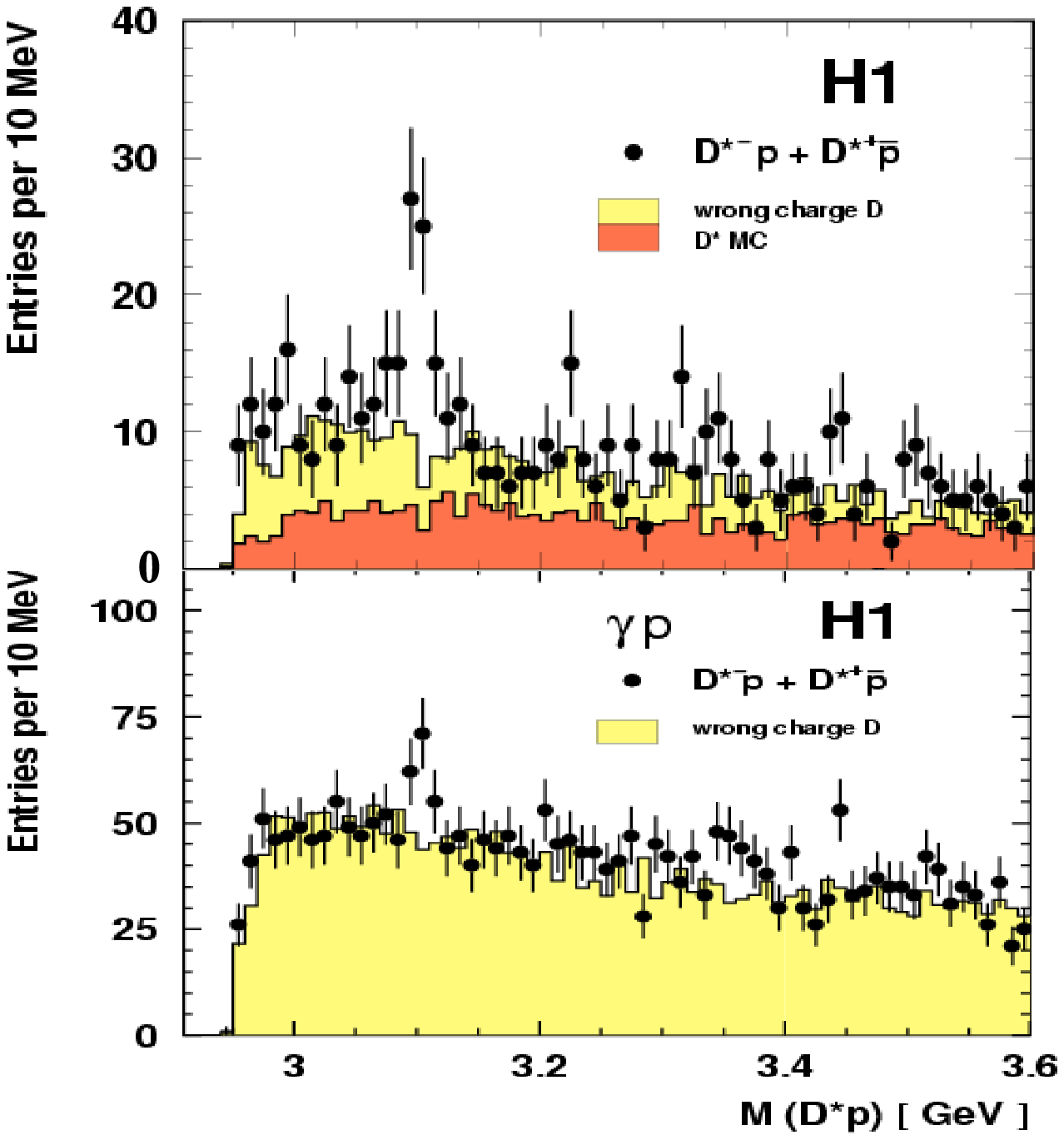}
\caption{Invariant mass of the opposite signs $D^*p$ combinations for
the DIS (top) and photoproduction (bottom) data samples of the H1 Collaboration.}
\label{fig:h1_cpq}
\vspace{-6mm}
\end{minipage}
\hspace{\fill}
\begin{minipage}[t]{75mm}
\includegraphics*[scale=0.56]{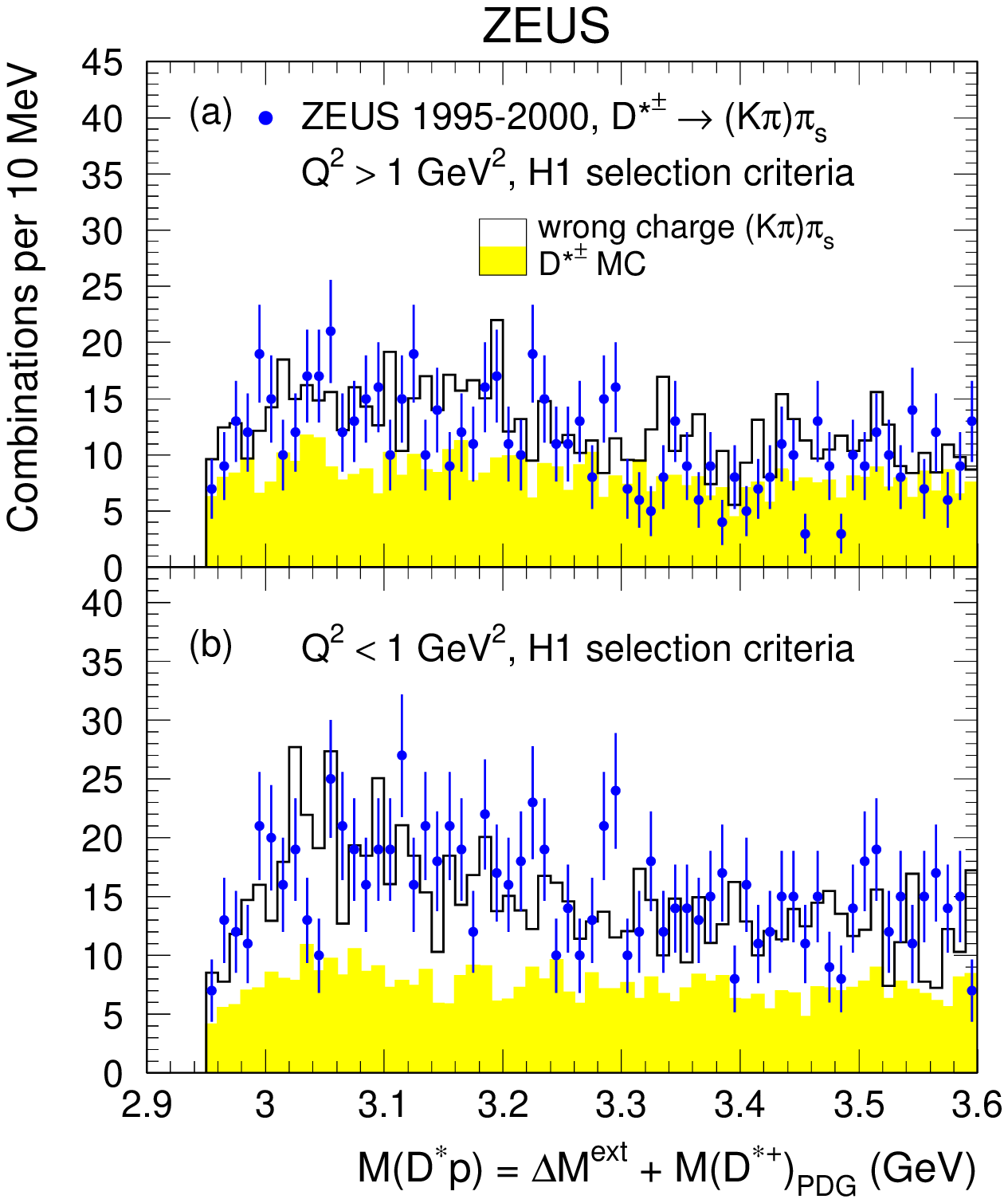}
\caption{Invariant mass of the opposite signs $D^*p$ combinations for
the DIS (top) and photoproduction (bottom) data samples of 
the ZEUS Collaboration with selection criteria similar to H1.}
\label{fig:zeus_cpq}
\vspace{-6mm}
\end{minipage}
\end{figure}

The existence of the strange pentaquark $\theta^{+}=uudd\bar{s}$ implies
that a pentaquark with the charmed quark, $uudd\bar{c}$, could also
exist. If the mass of such a state is above the sum of the mass of a
$D^*$ and a proton, it may decay to $D^{*\pm}p^{\mp}$. The H1 Collaboration
found~\cite{h1_cpq} a narrow signal in the $D^{*\pm}p^{\mp}$ invariant mass
at 3.1 GeV with a width consistent with the detector resolution. The $D^*$ 
mesons were identified using the 
$D^{*\pm} \to D^0\pi^{\pm} \to (K^{\mp}\pi^{\pm})\pi^{\pm}$ decay channel.
A clean signal is seen in DIS ($Q^2 > 1.0$ GeV$^2$) and, independently,
in the photoproduction sample ($Q^2 < 1.0$ GeV$^2$), as shown in 
figure~\ref{fig:h1_cpq}. 
The probability of the background to fluctuate to the
signal in the DIS sample was estimated to be less than $4\cdot10^{-8}$ which 
corresponds to
$5.4\sigma$ in terms of Gaussian standard deviations.
The ZEUS Collaboration made a similar search for such a state~\cite{zeus_cpq}. 
The ZEUS $D^*$ mesons were reconstructed using two decay channels of 
the $D^0$, 
$D^0 \to K^{\mp}\pi^{\pm}$ and $D^0 \to K^{\mp}\pi^{\pm}\pi^+\pi^-$.
The integrated luminosity of the ZEUS data sample is 1.7 times
larger than that of the H1 data sample. 
No narrow signal was observed in either of the $D^0$-decay modes. 
The selection criteria
were different in the two analyses,
resulting in different phase space region explored and in a larger yield
of reconstructed $D^*$ in the ZEUS investigation.
The ZEUS Collaboration presented also results in the region similar to the H1 
study and found no indication of a narrow resonance, 
neither in the DIS nor in the photoproduction sample 
(see figure~\ref{fig:zeus_cpq}).
New data from HERA-II will help to resolve this discrepancy in the
results of the two experiments.

It is a pleasure to thank the organizers for the warm and joyful atmosphere
in a most interesting and remarkably well prepared conference. I also wish
to thank my colleagues in H1 and ZEUS, for providing the data and results
presented in this report and for all their help given to me.

\end{document}